# Electro-Optic Search for Critical Divergence of the Charge-Density-Wave Diffusion Constant at the Onset of Depinning

L. Ladino, E.G. Bittle, M. Uddin, and J.W. Brill

Dept. of Physics and Astronomy, University of Kentucky, Lexington, KY

40506-0055

**Abstract:**
We have used electro-reflectance measurements to study charge-density-wave (CDW) polarization dynamics at voltages near the CDW depinning onset ($V_{on}$) in the quasi-one-dimensional conductor $K_{0.3}MoO_3$ ("blue bronze"). For low voltages, where the phase-slip rate is low, it is expected that the polarization relaxation time should be inversely proportional to the CDW diffusion constant, which is expected to diverge at $V_{on}$. At T = 78 K, we observe saturation of the relaxation time at low voltages, suggesting that we are in the low phase-slip limit and allowing us to estimate the non-critical value of the CDW diffusion constant $D(\infty) \sim 0.02 \text{cm}^2/\text{s}$, consistent with the measured phason velocity. At other temperatures, the relaxation time increases with decreasing voltage even at the lowest voltages we could measure. In no case do we observe the expected "critical speeding up", setting an upper limit on the critical region of $|V/V_{on} -1|_{critical} < 0.06$.

PACS#: 71.45.Lr, 78.20.Jq, 72.15.Nj, 64.60.Ht

For over three decades, there has been extensive research in sliding charge-density-waves (CDWs) in quasi-one-dimensional conductors as prototype systems for study of the effects of quenched disorder on an elastic medium.[1,2] For low applied forces, the CDW is pinned to the lattice by impurities, but at large voltages it can break free and carry collective current. One of the central questions has been the extent to which the onset of CDW sliding, at applied threshold voltage $V_{on}$, can be considered a dynamic critical transition of an elastic medium, as first proposed by Fisher.[3] From renormalization group (RG) calculations, the CDW drift velocity is expected to vary as $v_{CDW} \alpha \ [V/V_{on} -1]^{\zeta}$, with critical exponent $\zeta \sim 5/6$ (in three-dimensions),[4] but experimental attempts to measure the critical exponent of CDW velocity (i.e. current) have yielded a variety of values for $\zeta$,[5] possibly reflecting effects of imperfect electrical contacts on the CDW velocity. Indeed, the electric field and CDW current are non-uniform, increasing and decreasing respectively near (typically ~ 100 μm) current contacts where single particle current must convert to collective current,[6,7] so that interpreting I-V measurements is complicated even for non-perturbative contacts.[5] Furthermore, there have been no estimates of the width of the critical region for sliding CDWs; below threshold, the critical region has been theoretically estimated[8] to be $|V/V_{on} -1| < 0.1$, but, as discussed in Ref. (8), there is no obvious relationship between critical behavior above and below $V_{on}$.

In addition to the CDW velocity, other properties are expected to have critical behavior. The RG calculations[4] have shown that the dynamic phase (i.e. velocity) correlation length diverges as $\xi \alpha \ |V/V_{on} -1|^{-\nu}$, with $\nu \sim 1/2$ (different from the exponent for the static correlation length below threshold[8]), while changes in the CDW phase diffuse with *weakly divergent* diffusion constant $D \alpha \ |V/V_{on} -1|^{-2\nu+\zeta}$. However, we know of no experimental measurements of the critical properties of D or $\xi$.

When a voltage above threshold is applied to a CDW conductor, the CDW compresses on one side of the sample and rarefies on the other, resulting in a spatially dependent CDW phase, $\phi$ (or, equivalently, shift in CDW wavevector $\Delta Q \equiv \partial\phi/\partial x$). These CDW strains have been observed in $NbSe_3$[1] with transport[6] and x-ray diffraction[7] measurements and in blue bronze[1,9,10] ($K_{0.3}MoO_3$) using an electro-optical technique.[11] Changes in phase have been found to obey the diffusion-like equation:[6,12]

$$\partial\phi/\partial t \approx \beta(I-V_{on}/R_0) + D \ \partial^2\phi/\partial x^2 - \int r_{ps} dx \qquad (1),$$

where I is the total current in the sample, $R_0$ is the Ohmic resistance below threshold, $\beta$ is a constant, the diffusion constant is proportional to the CDW elasticity (see below), and the phase slip rate $r_{ps}$ is a strongly non-linear function of strain ($\partial\phi/\partial x$). Phase-slip constitutes an irreversible, i.e. "plastic", change in the local phase required for current conversion.[6] Because of the phase-slip term, the response time ($\tau$) for phase changes decreases rapidly with increasing current,[12] but for voltages near threshold $r_{ps}$ will be very small, giving

$$\tau(V) = L^2/\pi^2 D(V), \quad (2)$$

where L is the sample length and D is expected to be constant except in the critical region. Then one would expect the time constant to actually slowly *decrease* in the critical region as D diverges: $\tau \alpha |V/V_{on} -1|^{2\nu-\zeta} \sim |V/V_{on} -1|^{1/6}$. Indeed, the RG calculations of Ref. [3,4] assume that this "elastic" (i.e. negligible phase-slip) condition holds in the critical regime near threshold. (Alternatively, it has been suggested that significant amounts of phase-slip occur at *all* voltages and *throughout* the sample, so that depinning becomes a rounded first order transition,[13] and some experiments do suggest that phase slip is pervasive.[14] However, the measured spatial profiles of CDW strain[6,7,11] certainly indicate that phase-slip concentrates at the contacts and grows rapidly with voltage above threshold. While our present results will be discussed assuming that phase slip vanishes at threshold, the validity of this assumption remains an open question.)

In recent work on blue bronze,[15] we measured the voltage dependence, for $V > 1.5 V_{on}$, of the relaxation time by applying symmetric square-wave voltages at different frequencies, $\omega$, to the sample and measuring the relative change in reflectance ($\Delta R/R$) or transmittance ($\Delta\theta/\theta$) in phase and in quadrature with the applied square-wave. The electro-optic responses, assumed to be proportional to the CDW strain, were fit to a modified harmonic oscillator expression, e.g.:

$$\Delta R / R = A_0 / [1 - \omega^2/\omega_0^2 + (-i\omega\tau_0)^\gamma]. \qquad (3)$$

All four fitting parameters depend on sample, voltage, and position (i.e. distance from a current contact). At higher voltages, $\gamma=1$, but it generally was observed to decrease at smaller voltages (e.g. see Figure 3b, below), corresponding to a distribution of relaxation times. The average relaxation time was found to increase with decreasing voltage as $\tau_0 \sim V^{-p}$, with $p \sim 3/2 \pm 1/2$. (The electro-optic relaxation time is over three orders of magnitude larger than the dielectric relaxation time,[16] reflecting the fact that dielectric measurements probe local oscillations of the CDW phase rather than diffusion over macroscopic length scales.) The resonance term in Eqtn. (3) corresponds to a delay in the electro-optic response, with delay time $1/\omega_0$ tending to vanish at points adjacent to the current contacts, which we interpreted in terms of flow of the CDW strain from the contacts.[12,15] (Finite values of $\omega_0$ adjacent to the contact, generally most noticeable at high voltages, would then reflect our finite spatial resolution.) Finally, in some cases, the quadrature response was observed to become inverted at low-frequencies, an effect not included in Eqtn. (3) and corresponding to a decay of CDW strain at long times.

In the present work, we have attempted to observe critical behavior in the electro-optic relaxation time, and hence diffusion constant, by extending these frequency dependent measurements in blue bronze closer to threshold: $V < 1.1 V_{on}$. At one temperature we observe saturation of $\tau_0$ at low voltages, allowing us to estimate the "non-critical" value of the CDW diffusion constant. However, we do not observe the "critical speeding up" expected from the divergence of D in Eqtn. (2), giving an upper limit to the critical voltage above threshold. (Note that the measured time constants are orders of magnitude too *short* to be associated with the very slow CDW drift velocity.[15])

The experimental technique is similar to that used in Ref. (14). A blue bronze crystal, cleaved to a thickness between 5 and 10 μm, was chosen to have similar voltage and frequency-dependences of its electro-transmittance and electro-reflectance, showing that the CDW strains were fairly uniform in the sample cross-section,[15] well-defined onset voltages, and well-defined relaxation peaks, i.e. relatively large values of γ and negligible delays (near the contacts) and low-frequency decays at low voltages. Gold films were evaporated on the ends of the crystal to serve as current contacts. The resulting sample length and width were 650 μm and 120 μm. The sample was cooled in an optical cryostat and light (ν ~ 850 cm$^{-1}$, polarization perpendicular to the high conductivity axis of the sample, power < 100 μW) from a tunable infrared diode laser focused on a small (~ 50 x 50 μm$^2$) spot on the sample using an IR microscope. A symmetric, bipolar square-wave voltage, 1.5 Hz < ω/2π < 2 kHz, was applied to the sample, periodically reversing the CDW strain, and the resulting relative change in reflectance or transmittance measured. All measurements were made at relatively high temperatures where the CDW is believed to be weakly (i.e. collectively) pinned[5] and the critical models[3,4] are relevant, with the light spot adjacent to a current contact to minimize the resonance term in Eqtn. (3), as discussed below.

Figure 1 shows the voltage dependence of the electro-reflectance, both in phase and in quadrature with a ω/2π = 25 Hz square-wave at T = 78 K. Also shown is the voltage dependence of the dc resistance of the sample. Note that the electro-optic onset voltage (determined with a precision of ~ 0.3 mV), where the CDW becomes depinned, is below the threshold for non-linear current, $V_T$;[11,17] the difference is approximately the voltage needed for phase-slip.[17]

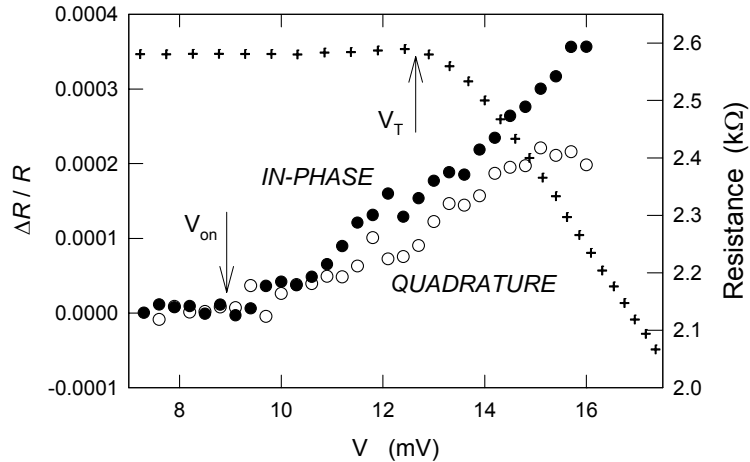

**Figure 1.** Voltage dependence of the electro-reflectance at T = 78 K. The responses both in phase (closed symbols) and in quadrature (open symbols) to the applied 25 Hz square wave are shown. Also shown is the voltage dependence of the dc resistance (crosses).

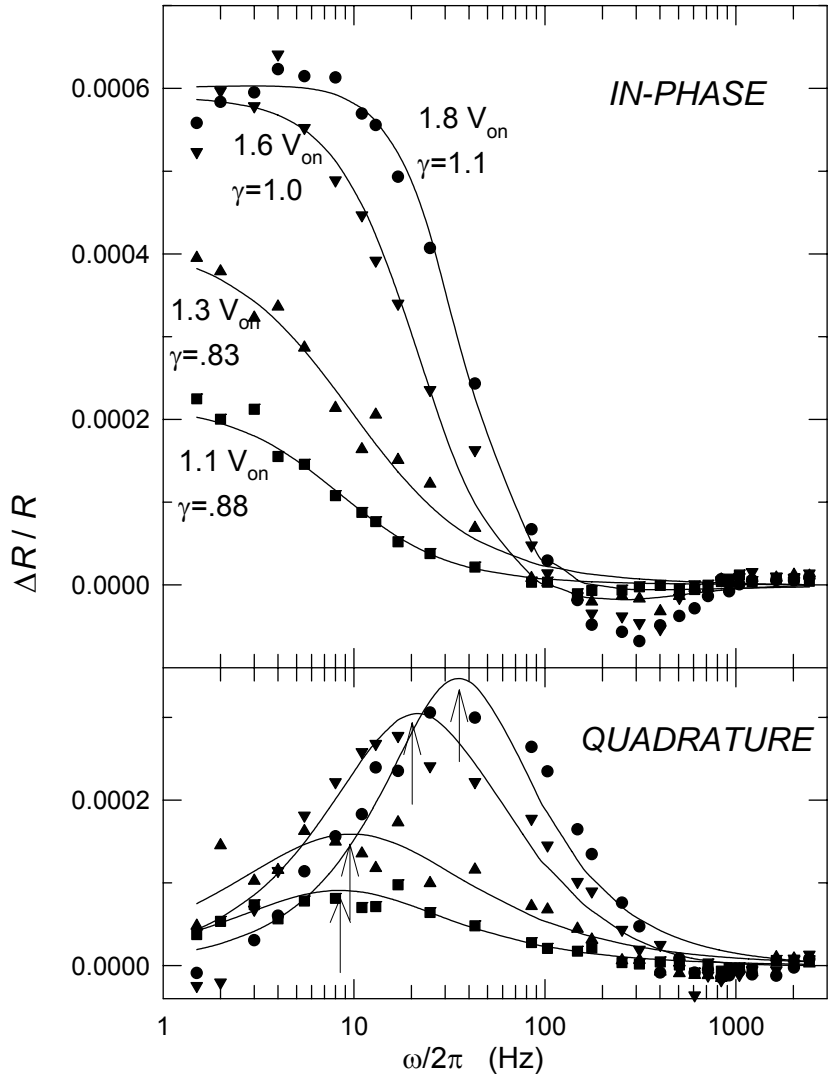

**Figure 2**. Frequency dependence of the electro-reflectance at T = 78 K. The in-phase and quadrature responses at a few voltages are shown. The curves show the fits to Eqtn. (3) with the resonance terms omitted (i.e. $\omega_0 = \infty$); the values for $\gamma$ for each fit are given. The vertical arrows show the values of $1/2\pi\tau_0$ of the fits.

Figure 2 shows the frequency dependence of the electro-reflectance at T = 78 K at a few voltages. Note that at the higher voltages the in-phase response becomes inverted at high frequency, corresponding to the resonance term in Eqtn. (3), but that this inversion is negligible at the lower voltages, so these fits are not sensitive to $\omega_0$ (i.e. $\omega_0/2\pi > 1$ kHz). Therefore, for internal consistency, fits were made to Eqtn. (3) *omitting the resonance term* at all voltages; the fitting curves are shown in the Figure. While these do a poor job at the higher frequencies for the higher voltage curves, including the resonance term mostly has the effect of decreasing the value of the exponent ($\gamma$) and has no significant effect on $\tau_0$. Indeed, the fitted values of $1/\tau_0$, shown by the vertical arrows in Figure 2, agree closely with $\omega_{peak}$, the peak frequency of the quadrature response. This is true even for $\gamma \sim 0.7$, our smallest value (see Figure 3b), where the resulting distribution of relaxation times is a decade wide.[18] Hence our values of $\tau_0$ are largely independent of the fitting function.

Figure 3 shows the dependence of the fitting parameters on voltage, normalized to $V_{on}$ and plotted on *a linear scale,* at a few temperatures. Except at T = 78 K, the relaxation time (Figure 3a) is still increasing at the lowest voltage at which, because of the rapidly decreasing amplitude of the response near $V_{on}$ (Figure 3c), we could determine it. Note that the maximum value of the time constant is $\tau_0(max) \sim 20$ ms so that $1/2\pi\tau_0(max) \sim 8$ Hz is within our measured frequency window and we are not instrumentally limited. Also note that the voltage dependence of $\tau_0$ is considerably stronger than for the samples previously measured;[15] this sample dependence of CDW strain properties is presumably a manifestation of different contact qualities as well as distributions of pinning impurities and other defects.[11,15]

As mentioned above, the decrease in $\tau_0$ with increasing current at "high voltages" reflects the increasing rate of phase slip. It is surprising that this behavior continues to voltages below $V_T$, the threshold of measurable non-linear current (shown by vertical arrows in the figure). That is, the polarization dynamics are still dominated by "plastic" phase-slip at voltages as low as $V \sim 1.06 V_{on}$. While the resulting strong voltage dependence of $\tau_0$ could mask any weak divergence of D, the fact that phase-slip dominates indicates that we are not yet in the elastic, critical regime considered in Reference (4).

At T = 78 K, on the other hand, the relaxation time saturates at low voltage (also apparent in Figure 2), so here we are presumably in the elastic and diffusive limit. This is consistent with the fact that here $V_T/V_{on} \sim 1.4$, while $V_T/V_{on} < 1.2$ at the other temperatures, indicating a larger voltage region in which the rate of phase-slip is very low at 78 K than elsewhere. However, even at 78 K there is no indication of a critical decrease in relaxation time at the lowest voltages, again giving $|V/V_{on} -1|_{critical} < 0.06$ as an upper limit of the critical voltage region; this is the essential result of this work.

We can also use Eqtn. (2) with the saturation value of the relaxation time, $\tau_0(sat) \sim 20$ ms to estimate the non-critical value of the diffusion constant:[19] $D(\infty) \sim 0.02$ cm$^2$/s. Then taking[6]

$$D(\infty) = (Q/en)^2 K/(\rho_0+\rho_{CDW}), \quad (4)$$

where $\rho_0$ and $\rho_{CDW}$ are the single-particle and (high-field) CDW resistivities, with $\rho_{CDW} \ll \rho_0$ ($\sim 0.6$ $\Omega$·cm at 80 K[9,10]) in semiconducting blue bronze, n $\sim 5 \times 10^{21}$ cm$^{-3}$ is the condensate density,[10] Q $\sim 0.7$ Å$^{-1}$,[10,20] and K is the CDW elastic constant, we find that *K* $\sim$ 0.1 eV/Å. This value is an order of magnitude larger than that estimated for NbSe$_3$,[6,21] which is appropriate since *K* is expected to increase with decreasing temperature in semiconducting blue bronze as the density of screening quasiparticles falls, whereas NbSe$_3$ remains metallic in the CDW state.[1] This increased stiffening in blue bronze has also been observed as a large increase in phason velocity, $v_\phi$, above its mean-field value at low temperatures.[20] In fact, our value of $K$ is consistent with that expected from the (T=80 K) value of $v_\phi \sim 1.4 \times 10^6$ cm/s;[20] i.e.

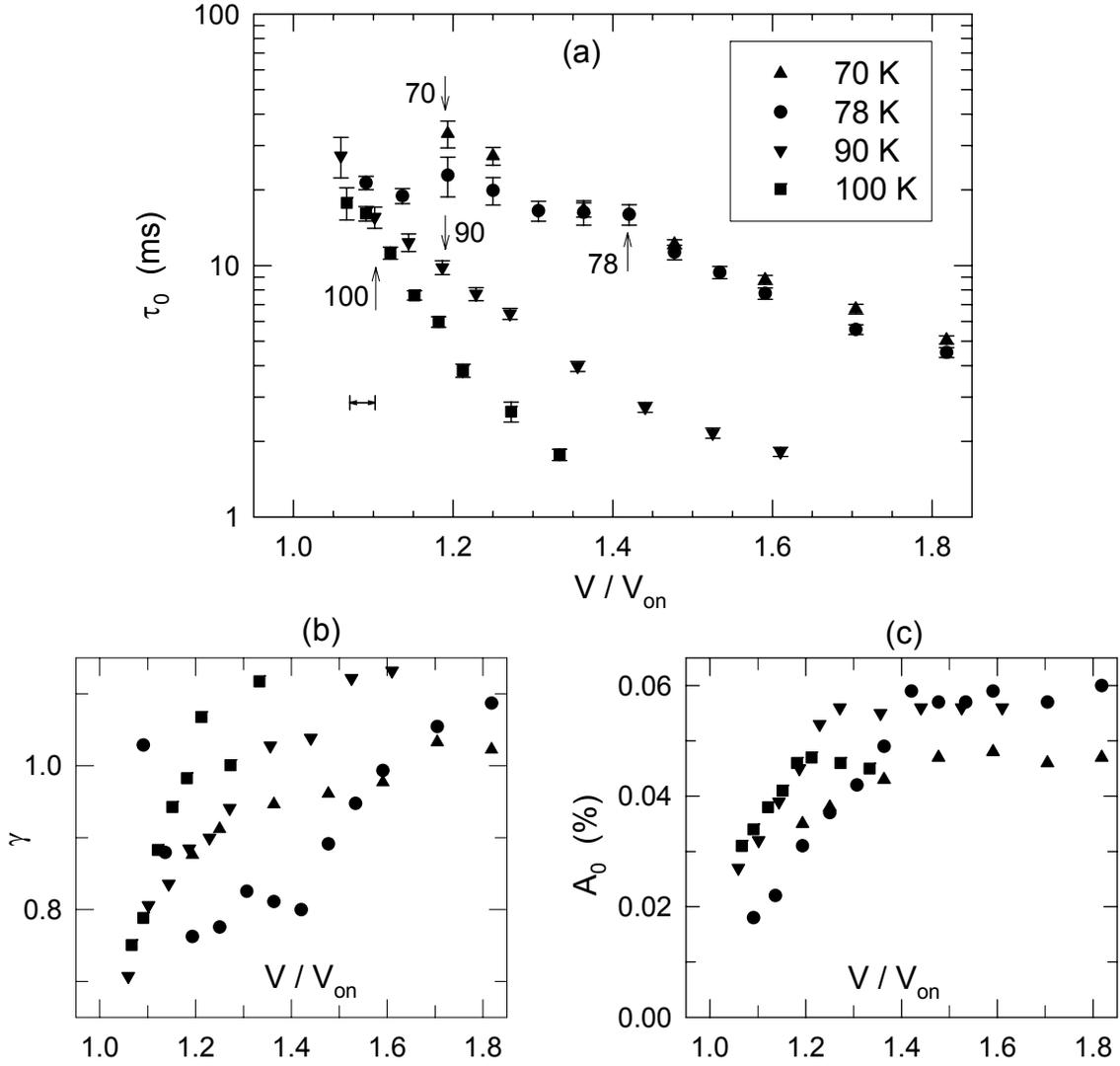

**Figure 3.** Voltage dependence of the fitting parameters [(a) $\tau_0$, (b) $\gamma$, and (c) $A_0$] at the temperatures indicated. In (a), the fitting uncertainties for $\tau_0$ are indicated, the horizontal bar shows the uncertainty in $V/V_{on}$ at low voltage, and the vertical arrows show the values of $V_T/V_{on}$ at each temperature.

$$K = (2 Q / \pi) \, m_F \, v_\phi^2, \quad (5)$$

where the factor in parenthesis is the linear density of CDW electrons (two bands condense into the CDW state giving 4 electrons/wavelength[10,22]) and the effective "Fröhlich" mass[1] has been estimated as $m_F \sim (150\text{-}350)\, m_e$.[20]

Alternatively, one might have suspected that as phase-slip becomes negligible near threshold, diffusion doesn't occur across the whole sample length but becomes limited to the correlation length. Then instead of Eqtn. (2), one would have $\tau \approx \xi^2/D$, and the relaxation time would critically *diverge* with exponent $-\zeta = -5/6$.[4] While our $T \neq 78K$ data are not inconsistent with such critical slowing down, we have presumably not reached the elastic limit appropriate to the RG calculations[4] at these temperatures. On the other hand, the 78 K saturated value of $\tau_0(sat) \sim 20$ ms, which we do associate with elastic behavior, seems too long to be due to diffusion across a non-critical value of $\xi$. Attempts to measure the correlation length in blue bronze with x-ray diffraction have been resolution limited so it is only known that $\xi > 1.6$ μm,[23] but it is unlikely that the non-critical correlation length is large enough (e.g. > 100 μm) to give a physical value for *K* from Eqtn. (4).

In principal, a straightforward test of $\tau_0(sat) = L^2/\pi^2 D(\infty)$ would be to compare saturated time constants for samples of different lengths. However, for low voltages, the "elastic" strain (and hence electro-optic response) $\sim 1/L$,[7] so working with much longer samples is not feasible. Working with much shorter samples would require preparing electrical contacts to ensure uniform current flow as well as improving our $\sim 50$ μm spatial resolution.

In conclusion, we have used an electro-optic technique to measure the CDW phase relaxation time as functions of voltage with voltages as small as $1.06\, V_{on}$. At most temperatures, the relaxation time is still increasing at the lowest voltages, indicating that the sample is still in the "plastic" (and presumably non-critical) limit. However, at one temperature we observe the relaxation time to saturate, with no indication of critical "speeding up" (i.e. divergence of the diffusion constant), giving an upper limit to the critical region of $|V/V_{on} - 1|_{critical} < 0.06$. The saturated relaxation time allows us to estimate the non-critical value of the CDW diffusion constant as $D(\infty) \sim 0.02$ cm$^2$/s, consistent with the value expected from the measured phason dispersion.[20]

**Acknowledgements:** We thank O. Narayan for very helpful discussions and R.E. Thorne for providing crystals. This work was supported by the National Science Foundation, Grant No. DMR-0400938.